\documentclass[%
 reprint,
nofootinbib,
 amsmath,amssymb,
 aps,
]{revtex4-1}

\usepackage{lipsum}
\usepackage{balance}
\usepackage{hyperref}
\usepackage{appendix}
\usepackage{enumitem}
\usepackage[user,titleref]{zref}
\usepackage[nameinlink,capitalize]{cleveref}
\usepackage{braket}
\usepackage{natbib}
\usepackage{subfigure}
\usepackage{tikz}
\usepackage{graphicx}
\usepackage{dcolumn}
\usepackage{bm}

\begin{document}

\preprint{APS/123-QED}

\title{Lensing of Gravitational Waves as a Novel Probe of Graviton Mass}

\author{Adrian Ka-Wai Chung}
\email{ka-wai.chung@ligo.org}
\affiliation{%
 Theoretical Particle Physics and Cosmology Group, Department of Physics, King’s College London, University of London, Strand, London, WC2R 2LS, United Kingdom
}%

\author{Tjonnie G.F. Li}
\email{tgfli@cuhk.edu.hk}
\affiliation{%
 Department of Physics, The Chinese University of Hong Kong, Shatin, N.T., Hong Kong
}%
\affiliation{%
Institute for Theoretical Physics, KU Leuven, Celestijnenlaan 200D, B-3001 Leuven, Belgium
}%
\affiliation{%
Department of Electrical Engineering (ESAT), KU Leuven, Kasteelpark Arenberg 10, B-3001 Leuven, Belgium
}%

\begin{abstract}
The diffraction patterns of lensed gravitational waves encode information about their propagation speeds.
If gravitons have mass, the dispersion relation and speed of gravitational waves will be affected in a frequency-dependent manner, which would leave traces in the diffraction pattern if the waves are lensed. 
In this paper, we study how the alternative dispersion relation induced by massive gravitons affects gravitational waves lensed by point-mass lenses, such as intermediate-mass black holes.
We find that the waveform morphology of lensed dispersive gravitational waves depends on the graviton mass more sensitively than their unlensed counterpart. 
Together with lensing amplification, the waveform-morphology modifications due to lensing can improve the measurement accuracy of the graviton mass.
A single lensed gravitational-wave signal enables us to measure the graviton mass with an accuracy comparable with the combined measurement across $\mathcal{O}(10^3)$ unlensed signals.
Our method allows us to incorporate lensed gravitational-wave signals into existing graviton-mass measurements. 
Our method can also be generalized to other lens types, gravitational-wave sources, and detector networks, preparing ourselves for thoroughly understanding the nature of gravitational waves in the era of lensed gravitational-wave astronomy. 
\end{abstract}

\pacs{Valid PACS appear here}
\maketitle


\section{Introduction}
When gravitational waves (GWs) propagate near a massive compact object, the propagation direction will be changed due to the gravity of the compact object \cite{Lensing_GW_01, Lensing_GW_02, Lensing_GW_03, Lensing_IMBH_02, Takahashi_2003, Takahashi_2017}. 
This phenomenon is known as lensing. 
The lensed waves interfere among themselves to form new wave patterns. 
The lensing pattern depends on the nature of the lens, lensing geometry, and interference between different (possible) rays.
Lensing is a crucial astrophysical probe. 
For example, lensing of GWs might provide us with the information of the existence of intermediate-mass black holes (IMBHs) (of mass $ \sim 10^2 M_{\odot} - 10^3 M_{\odot}$) \cite{Lensing_IMBH_01}, cosmological expansion \cite{Hubble_constant, Cremonese2021} and testing general relativity \cite{Suvodip01, Suvodip02, Collett_Bacon_2017, Fan_2017, Baker}.
Since the direct detection of GWs, searches for lensed GWs have been popular \cite{Lensing_O1_O2, O3a_lensing_paper, Lensing_LIGO_01}. 
Nonetheless, no conclusive evidence of lensed GW signals has been found as of the time of writing \cite{O3a_lensing_paper}. 

Another popular aspect of studies is the dispersion relation of GWs \cite{LIGO_01, LIGO_02, LIGO_03, LIGO_04, LIGO_05, LIGO_06, LIGO_07, LIGO_08, LIGO_09, LIGO_10, LIGO_11, LIGO_12, LIGO_13, LIGO_14, LIGO_15}.
According to general relativity, in vacuum, GWs propagate at the speed of light and obey the dispersion relation of $ \omega = c k $, where $ \omega $ is the angular frequency, $c$ is the speed of light, and $ k $ is the wave number. 
This dispersion relation implies that gravitons are massless. 
On the other hand, some alternative theories explore the possibility of massive gravitons, see, e.g., \cite{MG_review_01} for a review. 
If the gravitons have mass, GWs obey an alternative dispersion relation, leading to different propagation speeds for various frequencies. 
As GWs propagate, dephasing will be developed across different frequencies.
By (not) measuring the dephasing, we can constrain the graviton mass \cite{m_g_bound_review_01, m_g_bound_review_02, Will_2012}. 
Alternatively, the near-field behavior of black holes has also been suggested as a probe of the graviton mass \cite{Chung_2019}. 
As of the time of writing, no evidence of massive gravitons has been found using this method \cite{LIGO_03, LIGO_06, LIGO_07, LIGO_11}. 

These methods are capable of extracting the graviton mass from unlensed GW signals. 
On the other hand, lensed GW signals are expected to be detected in the future \cite{Ken_2016, Lensing_LIGO_01}. 
To thoroughly understand the nature of GWs in different astrophysical scenarios, developing a test of the dispersion of lensed GWs becomes increasingly pressing.
Moreover, since the amplitude of lensed GWs shows more variations across the frequency than the amplitude and phase of unlensed waves \cite{Takahashi_2003, lensingGW}, the waveform morphology of lensed dispersive GWs may depend on the graviton mass more sensitively than the unlensed GWs.
Besides, the amplification introduced by lensing may contribute to an improved measurement accuracy of the graviton mass compared to the unlensed case.
Furthermore, the dispersion relation of GWs corresponding to the massive graviton also changes the time delay of waves of different frequencies in different directions, leading to additional features of the resultant lensing pattern. 
These considerations prompt us to explore measuring the graviton mass from lensed GW signals.
Measuring the graviton mass by lensing also makes relevant tests more complete in at least two ways. 
First, lensing involves the strength of gravity intermediately between the near and far fields.
Second, our work enables us to incorporate lensed GW signals into the measurement of the graviton mass, which has thus far focused only on unlensed signals, better preparing ourselves for the era of lensed GW astronomy. 

Throughout the paper, $ m_g $ is in the unit of $ c = \hbar = 1 $ (so $ h =2 \pi $), while the mass of compact objects (such as black hole and lens) are in the unit of $ c = G = 1 $. 

\section{Lensing Pattern of Gravitational Waves with Dispersion}

\subsection{Assumptions and approximations}

This work makes a few assumptions:
\begin{enumerate}[label=A.\arabic*,start=1]
    \item Following \cite{Will_2012, Ajith_2010, Yagi_2015}, we assume perfect screening of gravity due to the mass of graviton \cite{Yunes_2018}.
    In other words, general-relativistic limits are recovered at a length scale shorter than the Compton wavelength of the graviton. 
    This assumption implies that we will ignore the effects on the dynamics of binary black hole mergers due to the graviton mass. 
    In the context of lensing, this assumption implies that at a sufficiently far distance $r$ the Newtonian gravitational potential due to a black hole (point-mass lens) of mass $M$ is given by $ - \frac{M}{r}$. 
    \item We focus only on the effects on GW lensing due to the graviton mass.
    Other consequences of lensing, such as modifications on polarization \cite{Lensing_polarization} and phase shift \cite{Lensing_phase_shift_01, Lensing_birefringence}, will be omitted. 
    These are acceptable approximations because including these effects will include more contrasting features to graviton-mass measurement. 
    In this work, we focus on improvements on the constraint or measurement accuracy of the graviton mass by considering the combined effects of lensing and dephasing induced by the graviton mass.  
\end{enumerate}

\subsection{Method}

If gravitons have mass, phenomenologically, the dispersion relation of GWs will be altered to \cite{Will_2012} 
\begin{equation}\label{eq:LIV_relation}
\omega^2 = k^2 + m_g^2, 
\end{equation}
where $m_g $ is the mass of graviton  \footnote{Alternatively, this equation can be interpreted as a definition of the massive graviton which leads to the dispersion of gravitational perturbations. 
In this work, we refer ``the mass of graviton" to $m_g$ defined by \cref{eq:LIV_relation}}. 
If $ m_g \ll k $, the propagation speed of dispersive GWs that obey this dispersion relation can be approximated by the following equation: 
\begin{equation}\label{eq:v_f}
v_g(f) \approx 1 - \frac{1}{8 \pi^2}\frac{m_g^2}{f^2}. 
\end{equation}
When propagating in a flat space-time, the dispersive GWs obeying \cref{eq:LIV_relation} will acquire a dephasing due to the difference in propagation speeds among different frequencies \cite{Will_2012}, 
\begin{equation} \label{eq:LIV_dephasing}
\begin{split}
& \Psi_{\rm disp}(f; m_g) = -\frac{\pi D_{0}}{\lambda_{g}^{2}} \frac{1}{(1+z)f}, 
\end{split}
\end{equation}
where $ \lambda_g = 1/m_g$ is Compton's wavelength of the graviton, $ D_0 $ is the propagation distance from the source to the detector and $ z $ is the redshift of the source binary. 
Thus, in the frequency domain, the waveform of unlensed dispersive GWs is 
\begin{equation}\label{eq:unlensed_waveform}
\tilde{h}_{\text{\rm disp}} (f) = \tilde{h}(f) e^{i \Psi_{\text{disp}}(f)},
\end{equation}
where $\tilde{h}(f)$ is the original (unlensed) GR waveform (see, e.g., \citep{Sathyaprakash_2009, IMRPhenom_2011, IMRPhenom_2014} for GR waveform approximants).

When encountering a massive compact object, such as an intermediate-mass black hole, GWs will be lensed. 
The lensing effect is characterized by the amplification function (or transmission factor) 
\citep{Schneider_01, Schneider_02}, $F$, which is the ratio of lensed-wave amplitude to unlensed-wave amplitude, 
\begin{equation}
\tilde{h}_{L}(f) = F(f) \tilde{h}(f), 
\end{equation}
where $ \tilde{h}_{L}(f) $ is the lensed waveform and $\tilde{h}(f) $ is the unlensed waveform. 
Given a lensing geometry, $ F(f) $ can be computed by \cite{Takahashi_2003, Takahashi_2017, Nakamura_02}
\begin{equation}\label{eq:Amp_fn_def}
\begin{split}
F(f; \vec{\theta}_s) = & \frac{D_L D_S}{D_{LS}}  \xi_0^2 \frac{(1+z_L)}{i} \frac{f}{v_g} \\
& \times \int d^2 \vec{\theta}_L \exp \left[2\pi i f t_d(\vec{\theta}_L, \vec{\theta}_s) \right], 
\end{split}
\end{equation}
where $v_g$ is GW propagation speed; $ D_L, D_S $, and $ D_{LS} $ are, respectively, the lens-to-observer distance, the source-to-observer distance, and the source-to-lens distance; $ z_L $ is the redshift of lens; $ \vec{\theta}_s$ is the displacement from optical axis to the source on source plane; $ \vec{\theta}_L$ is the displacement from optical axis to lens on lens plane; and $ t_d $ is the time delay between the lensed ray and unlensed ray,
\begin{equation}
t_d(\vec{\theta}, \vec{\theta}_s) = \frac{(1+z_L)}{v_g} \Bigg[ \frac{D_L D_S}{2 D_{LS}}|\vec{\theta}_s - \vec{\theta}|^2 - \psi(\vec{\theta}_s)\Bigg], 
\end{equation}
where $\psi(\vec{\theta}_{\rm s})$ is the lensing potential.
Overall, $t_d$ also depends on $v_g$, $ \vec{\theta}_s$, and lens $\vec{\theta}_L $, and $ \xi_0 $ is a length scale.

We note that the amplification function \cref{eq:Amp_fn_def} depends on $\frac{f}{v_g} $ as a whole. 
Thus, the amplification function of GWs of the massive graviton is just that of GWs without dispersion with the following replacement:
\begin{equation}
f \rightarrow \beta(f) f, 
\end{equation} 
where 
\begin{equation}\label{eq:refractive_index}
\beta(f) = \frac{c}{v_g (f)} \approx 1 + \frac{1}{2}\frac{m_g^2}{f^2}. 
\end{equation}
From \cref{eq:refractive_index}, we expect that the modifications to the lensing pattern due to the dispersion relation \cref{eq:LIV_relation} are manifest for $m_g \geq 10^{-14} \rm eV$, corresponding to the energy scale of $ hf $ at $ f= 10 \rm Hz$. 

As a proof of principle, in this work we focus on the case of a point-mass lens, such as a black hole. 
For a point-mass lens, the amplification function can be analytically evaluated as \cite{Takahashi_2003, Takahashi_2017}
\begin{equation}
\begin{split}\label{eq:Amp_fn_mg}
& F(f; M_{\rm len}, y, m_g) \\
& = \exp \Big(\frac{\pi}{4} w \beta \Big) \Big( \frac{w}{2} \beta \Big)^{i\frac{w}{2} \beta} \Gamma \Big( 1 - i\frac{w}{2} \beta\Big) {}_1 F_1 \Big( i\frac{w}{2} \beta, 1; i\frac{w}{2} \beta y^2 \Big),
\end{split}
\end{equation}
where $M_{\rm len} $ is the redshifted mass of the lens, $y$ is the impact parameter  of lensing, $\Gamma$ is the (complex) Gamma function, $ {}_1 F_1 $ is confluent hypergeometric function, and $ w = 8 \pi M_{\rm len} f$ is the dimensionless frequency. 
The resulting lensed waveform of GWs corresponding to the massive gravitons can be written as 
\begin{equation}\label{eq:lensed_waveform}
\tilde{h}_{L}(f; m_g) = F(f; M_{\rm len}, y, m_g) \tilde{h}(f) e^{ i \Psi_{\rm disp}(f)}. 
\end{equation}
Note that, according to \cref{eq:v_f}, GWs of different frequencies travel at different speeds. 
The only constant achromatic speed is the speed of light.
Therefore, the effects described by \cref{eq:lensed_waveform} are not degenerate with a constant change of propagation speed of GWs. 
Thus, the effects of the massive gravitons can be distinguished upon gravitational-wave detection.

\begin{figure}[t!]
  \centering
	\includegraphics[width=\columnwidth]{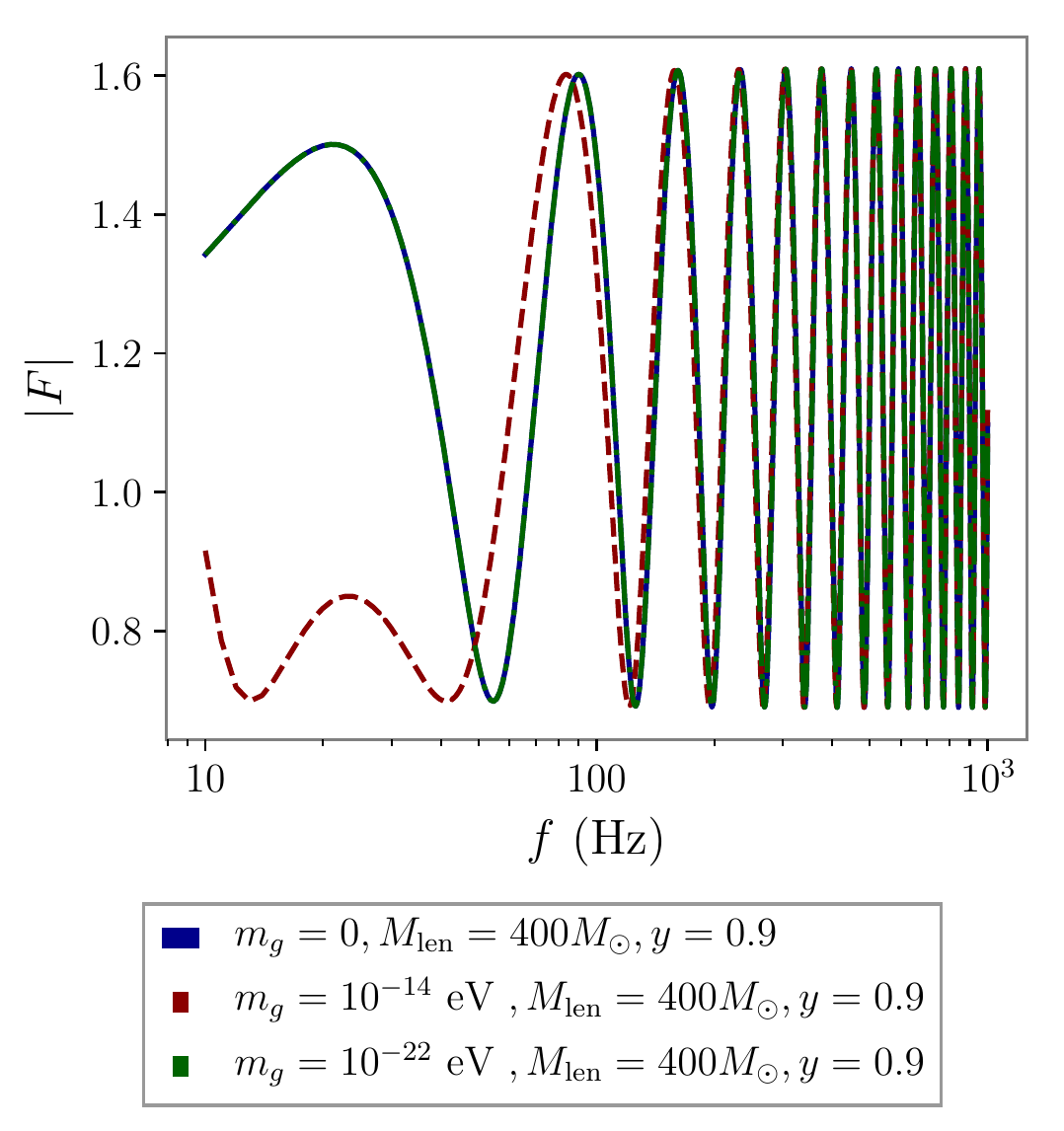}
	\caption{The lensing amplification function due to an intermediate mass black hole of redshifted mass $M_{\rm len}$ of 400 $ M_{\odot} $ at $ y = 0.9 $ corresponding to $m_g = 0$ (solid blue line), $ m_g = 10^{-14} \rm eV $ (dashed red line), and $m_g = 10^{-22} \rm eV$  (dashed dotted green). 
	At $m_g = 10^{-22} \rm eV $, which is close to the existing constraints by gravitational-wave detection, the change of the amplification due to $m_g$ is not very manifest as the solid blue line and dotted green line overlap almost entirely.}
  \label{fig:Amp_fn}
\end{figure}

\cref{fig:Amp_fn} plots the $ F(f) $ corresponding to the lensing by an intermediate-mass black hole of redshifted lens mass of $ 400 M_{\odot} $ and $ y = 0.9 $ for $ m_g = 0 $ (solid blue), $ m_g = 10^{-14} \rm eV $ (dashed red) and $m_g = 10^{-22} \rm eV $ (dotted green) as a function of $ f $. 
For $m_g = 10^{-14} \rm eV $, we find that the modifications of the amplification function is manifest for the low-frequency regime ($f \leq 10^2 \rm Hz$), in which $\beta$ changes significantly with $m_g$.
As GW frequency increases, the changes of the amplification function due to the alternative dispersion become increasingly less manifest because the ultrarelativistic limit $ E \approx p $ has been attained.
For $m_g = 10^{-22} \rm eV $, over $f\in [10, 10^3] \rm Hz$, the modifications due to the graviton mass are not visible, as expected because $|\beta(f)-1| \sim 10^{-15} \ll 1 $ for $m_g = 10^{-22} \rm eV $.

\cref{eq:lensed_waveform} suggests that GW lensing may help to improve the measurement of $m_g$ in at least three ways. 
\begin{enumerate}[label=R.\arabic*,start=1]
    \item \label{item:R1} Lensing changes the waveform morphology of the signal. 
    Specifically, because of the modulation by the amplification function (\cref{fig:Amp_fn}), the amplitude and phase of lensed GWs show more variations across the frequencies than the unlensed GWs.
    This beating pattern may make the waveform morphology of lensed dispersive GWs depend on the graviton mass more sensitively than unlensed waves. 
    \item \label{item:R2} Lensing increases the signal-to-noise ratio (SNR). 
    \item \label{item:R3} The graviton mass modifies the amplification function, making the waveform morphology of lensed dispersion GWs depend on the graviton mass even more sensitively.
\end{enumerate}
However, judging from \cref{fig:Amp_fn}, for $m_g $ close to the existing constraints of the graviton mass ($\sim 10^{-22} \rm eV $ \cite{LIGO_07, LIGO_11, LIGO_15}), the changes of the amplification function due to $m_g$ are not significant. 
Therefore, \ref{item:R3} is unlikely to contribute to any significant improvement. 
In what follows, we focus on investigating the roles of \ref{item:R1} and \ref{item:R2}. 

As a first step, we compare the similarity of the waveform of both lensed and unlensed dispersive GWs of a given $m_g$ to the dispersive waves of other $m_g$. 
In general, the similarity between two waveforms, $\tilde{h}_{1} (f)$ and $\tilde{h}_{2}(f)$, can be gauged by the match between $ \tilde{h}_{1} $ and $ \tilde{h}_{2} $, defined as
\begin{equation}
\mathcal{M} = \frac{\braket{\tilde{h}_{1}|\tilde{h}_{2}}}{\sqrt{\braket{\tilde{h}_{1}|\tilde{h}_{1}} \braket{\tilde{h}_{2}|\tilde{h}_{2}}}},
\end{equation}
where the braket notation denotes the noise-weighted inner product \cite{Inner_product}, 
\begin{equation}\label{eq:inner_product}
\braket{\tilde{h}_{1}|\tilde{h}_{2}} = 4 \text{Re} \int_{0}^{+\infty} df \frac{\tilde{h}_1 (f) \tilde{h}_2^{\dagger}(f)}{S_n(f)}, 
\end{equation}
and $ S_n (f) $ is the one-sided power-spectral density of the detector.
Throughout this work, we assume GW signals are detected by the Advanced LIGO and Virgo detectors operating at their design sensitivity \cite{LIGO_noise_curve, Virgo_noise_curve}. 
To investigate how sensitive lensed dispersive GWs depends on $m_g$, we chose 
\begin{equation}
\begin{split}
& \tilde{h}_{1}(f) = \tilde{h}_{L}(f; m_g = m_g^{\rm inj}), \\
& \tilde{h}_{2}(f) = \tilde{h}_{L}(f; m_g), 
\end{split}
\end{equation}
where $m_g^{\rm inj}$ is a given value of $m_g$ and $\tilde{h}_{L}$ is the lensed waveform defined by \cref{eq:lensed_waveform} and \cref{eq:Amp_fn_mg}. 
Using this waveform, we have defined a match as a function of $m_g$ for lensed dispersive GWs. 
Similarly, we can define a match for unlensed dispersive GWs by replacing $ \tilde{h}_{L}(f; m_g = m_g^{\rm inj}) \rightarrow  \tilde{h} (f; m_g = m_g^{\rm inj})$ and $ \tilde{h}_{L}(f; m_g) \rightarrow  \tilde{h} (f; m_g)$, where $\tilde{h} (f; m_g)$ is defined by \cref{eq:unlensed_waveform}. 
As $ \sqrt{\braket{\tilde{h}_{1}|\tilde{h}_{1}}}$ and $ \sqrt{\braket{\tilde{h}_{2}|\tilde{h}_{2}}}$ are, respectively, the SNRs of $\tilde{h}_{1}$ and $\tilde{h}_{2}$\footnote{Note that, throughout this work, SNR is defined with respect to the Advanced LIGO and Virgo detectors at the design sensitivity. }, $\mathcal{M}$ does not depend on the SNR of the waveform considered. 
Alternatively, $\mathcal{M}$ can be viewed as a normalized inner product between the two waveforms, and its magnitude is always smaller than 1. 
If $\tilde{h}_1$ and $\tilde{h}_2$ have more similarity, $\mathcal{M}$ is closer to unity. 
In particular, if $\tilde{h}_1(f) \propto \tilde{h}_2 (f)$, meaning that $\tilde{h}_1$ and $\tilde{h}_2$ have the same morphology, $\mathcal{M} = 1$. 

For the explicit calculations of $\mathcal{M}$, we consider: 
\begin{enumerate}[label=U.\arabic*,start=1]
    \item \label{item:U1} an unlensed waveform due to a GW150914-like source binary black hole \cite{LIGO_14} at a luminosity distance of 400 Mpc, whose SNR is 46, 
\end{enumerate}
\begin{enumerate}[label=L.\arabic*,start=1]
    \item \label{item:L1} a lensed waveform of the unlensed waveform by an IMBH of reshifted mass of $ 400 M_{\odot} $ and impact parameter $ y = 0.9 $, whose SNR is 57.
\end{enumerate}
This mass of lens is chosen because IMBHs of similar masses are hoped to be discovered by GW lensing \cite{Lensing_IMBH_01}. 
This value of $y$ is chosen because IMBH lensing is more likely to occur at a larger $y$ (see the subsequent discussion of the prior of $y$). 
The existing constraints on $m_g$ by GW detection \cite{LIGO_12, LIGO_15, LIGO_16} suggest that we can probe the existence of massive gravitons of $\sim 10^{-22}$ eV via GW detection.
Thus, we consider $m_g^{\rm inj} = 10^{-22} \rm eV $. 

\begin{figure}[tp!]
  \centering
	\includegraphics[width=\columnwidth]{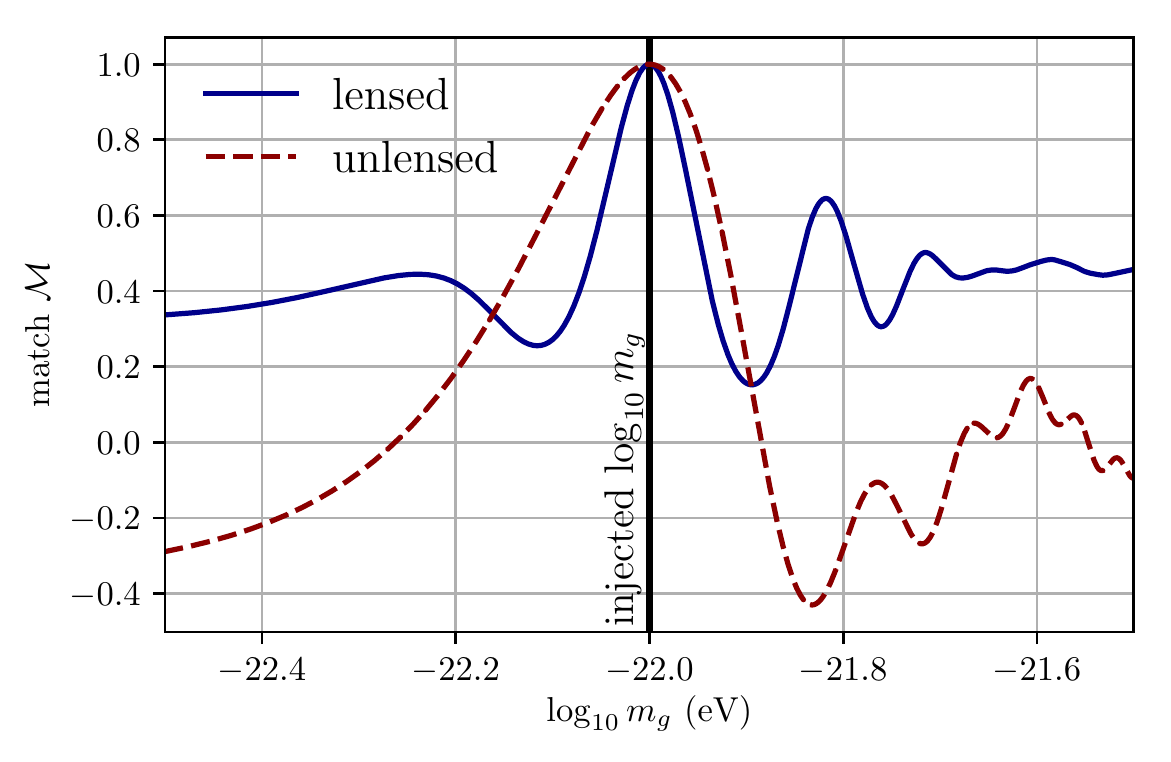}
	\caption{The match, a function gauging the similarity between waveforms, of lensed GWs and unlensed GWs as a function of the graviton mass. 
	In particular, we compare the similarity between lensed dispersive GWs of $m_g = 10^{-22} \rm eV$ to lensed dispersive GWs of other $m_g$  (solid blue) and the similarity between unlensed dispersive GWs of $m_g = 10^{-22} \rm eV$ to unlensed dispersive GWs of other $m_g$. 
	The match of lensed dispersive GWs shows a narrower peak, suggesting that the waveform morphology of lensed dispersive GWs vary more sensitively with $m_g$. 
	As we shall see, this character of lensed GWs can lead to a better measurement accuracy of $m_g$ over unlensed GWs. 
	}
  \label{fig:match}
\end{figure}
\cref{fig:match} shows the match $\mathcal{M}$ of the lensed waveform \ref{item:L1} (solid blue line) and the unlensed lensed waveform \ref{item:U1} (dashed red line) as a function of $m_g$. 
For both waveforms, $\mathcal{M}$ peaks at unity at the injected $m_g$; this is reasonable because at the injected $m_g$, $\tilde{h}_1(f) = \tilde{h}_2(f) $.
Nonetheless, the match of the lensed waveform peaks more sharply than that of the unlensed signal, suggesting that, at the same SNR, the morphology of lensed dispersive GWs vary more sensitively than the unlensed waves at $m_g \approx 10^{-22} \rm eV$. 
This feature hints that lensing may have advantages in measuring $m_g$ over unlensed waves. 
For $m_g$ far away from the injected $m_g$, $\mathcal{M}(m_g)$ drops well below unity for both the lensed and the unlensed waveform, because the waveform at those values of $m_g$ is significantly different from that at $m_g = 10^{-22} \rm eV$. 
At last, we note that $\mathcal{M}$ for both lensed and unlensed waves is oscillatory for $m_g$ far away from $m_g^{\rm inj}$. 
This is because $\tilde{h}_2 (f) \sim \tilde{h}_1 (f) e^{i \left[\Psi_{\rm disp}(f; m_g^{\rm  inj}) - \Psi_{\rm disp}(f; m_g) \right]} $, where $\Psi_{\rm disp}(f; m_g)$ is defined by \cref{eq:LIV_dephasing}, leading to 
\begin{equation}
\braket{\tilde{h}_{1}|\tilde{h}_{2}} \propto \int_{0}^{+\infty} df \frac{|\tilde{h}_1 (f)|^2}{S_n(f)} \cos \left\{ \frac{\pi D_0 \left[m_g^2 - \left( m_g^{\rm inj} \right)^2 \right]}{(1+z) f} \right\}. 
\end{equation}
The cosine term in the integrand makes $\mathcal{M}$ oscillatory over $m_g$. 

\section{Parameter Estimation}

\subsection{Mock signals}

To further investigate how lensing modifications of waveform morphology (\ref{item:R1}) and SNR (\ref{item:R2})  may help to improve the measurement of $m_g$, we analyze a mock signal of \ref{item:L1} and \ref{item:U1} that is injected into simulated Gaussian noises assuming the design sensitivity of the Advanced LIGO and Virgo detectors \cite{LIGO_noise_curve, Virgo_noise_curve}. 
We also inject
\begin{enumerate}[label=L.\arabic*,start=2]
    \item \label{item:L2} a lensed signal which is identical to \ref{item:L1} except the source binary is at 500 Mpc. 
\end{enumerate}
The luminosity distance of the source binary of \ref{item:L2} is increased so that the SNR of \ref{item:L2} is the same as that of \ref{item:U1}. 
On the other hand, it is estimated that the Advanced LIGO and Virgo detectors will detect 0.05 IMBH-lensed events per year (or 1 IMBH-lensed event per $\sim$ 20 years) \cite{Lensing_IMBH_01}.
At its design sensitivity, the Advanced LIGO and Virgo are expected to detect $\leq 360 $ unlensed events per year \cite{LIGO_detection_rate}. 
Thus, a more fair comparison will be with the posterior of $m_g$ combined across $20 \times 360 \sim $ 7000 unlensed signals. 
In practice, the combined measurement accuracy of $m_g$ will be dominated by the signal with the best measurement accuracy, which depends on the SNR of the signal \cite{m_g_bound_review_02}. 
Thus, we first simulated a population of $\sim 7000$ binary black-hole mergers according to \cite{LIGO_17}, each of which has an SNR of $\geq10$, approximately the minimum SNR for an event to be detectable by the Advanced LIGO and Virgo detectors \cite{LIGO_16, LIGO_15, LIGO_sensitivity}. 
Then, we inject the fourth signal, which is 
\begin{enumerate}[label=P.\arabic*,start=1]
    \item \label{item:P1} the unlensed signal that has the largest SNR (130) among the simulated 7000 unlensed events. 
\end{enumerate}
We represent the measurement of $m_g$ combined across these $ 7000$ simulated signals by the posterior of $m_g$ of \ref{item:P1}. 

\subsection{Bayesian inference}

We denote parameters describing the source binary by $\vec{\theta}_{\rm BBH} $ and parameters describing lensing by $\vec{\theta}_{\rm lens} = (M_{\rm len}, y)$. 
By Bayes' theorem, the posterior of $m_g$, $ \vec{\theta}_{\rm len} $ and $ \vec{\theta}_{\rm BBH} $ is given by
\begin{equation}\label{eq:Bayes_theorem}
\begin{split}
& p(\vec{\theta}_{\rm BBH}, \vec{\theta}_{\rm lens}, m_g|\tilde{d}, H, I) \\
& \propto p_{\rm BBH}(\vec{\theta}_{\rm BBH}|H, I) p_{\rm lens}(\vec{\theta}_{\rm lens}|H, I) p_{m}(m_g|H, I) \\
& \quad \times p( \tilde{d}|\vec{\theta}_{\rm BBH}, \vec{\theta}_{\rm lens}, m_g, H, I), 
\end{split}
\end{equation}
where $ p_{\rm BBH}(\vec{\theta}_{\rm BBH}|H, I) $, $p_{\rm lens}(\vec{\theta}_{\rm lens}|H, I)$ and $ p_{m}(m_g|H, I)$ are, respectively, the prior of $ \vec{\theta}_{\rm BBH} $, $ \vec{\theta}_{\rm lens} $ and $ m_g $, given the hypothesis $H$ that GWs may exhibit dispersion relation due to the massive gravitons and background information $I$, such as that the signal is lensed, the amplification function \cref{eq:Amp_fn_mg} and lensing geometry etc.
Since $ \vec{\theta}_{\rm BBH} $, $ \vec{\theta}_{\rm lens} $ and $ m_g $ should be independent, we have assumed that their priors are factorized.
$p( \tilde{d}|\vec{\theta}_{\rm BBH}, \vec{\theta}_{\rm lens}, m_g, H, I)$ is the likelihood that a binary black hole of $\vec{\theta}_{\rm BBH}$ and lens of $\vec{\theta}_{\rm len} $ will generate detected strain data $\tilde{d}_{D}$, 
\begin{equation}
\begin{split}\label{eq:likelihood}
& p(\tilde{d}|m_g, \vec{\theta}_{\rm len}, \vec{\theta}_{\rm BBH}, \vec{\theta}, H, I)  \propto \exp \left( - \frac{1}{2} \braket{\tilde{n}(f)|\tilde{n}(f)}\right), \\
& \tilde{n}(f)=\tilde{h}_{\rm D}(f; m_g, \vec{\theta}_{\rm len}, \vec{\theta}_{\rm BBH})  -\tilde{d}_{\rm D}, 
\end{split}
\end{equation}
where $\tilde{h}_{\rm D}(m_g, \vec{\theta}_{\rm len}, and \vec{\theta}_{\rm BBH})$ is the frequency-domain responses corresponding to detector $D$ by the waveform equation \cref{eq:lensed_waveform}.

Following \cite{Lensing_IMBH_01}, we place a uniform prior for $ M_{\rm len} $. 
For $ y $, we place a prior which is uniform for $ y^2 \in [0, 1]$ instead of $ y $. 
For $m_g$, we place a prior which is uniform for $\log_{10} m_g \in [-26, -20]$, covering the magnitude of the most updated constraints on $m_g$ \cite{LIGO_15} by GWs and for us to explore tighter constraints.
At last, the marginalized posterior of $ m_g $ can be obtained by marginalizing \cref{eq:Bayes_theorem} over  $ \vec{\theta}_{\rm BBH} $ and  $ \vec{\theta}_{\rm lens} $, 
\begin{equation}
\begin{split}
& p(m_g|\tilde{d}, H, I) \\
& = \int d \vec{\theta}_{\rm BBH} \int d \vec{\theta}_{\rm lens} p(\vec{\theta}_{\rm BBH}, \vec{\theta}_{\rm lens}, m_g|\tilde{d}, H, I). 
\end{split}
\end{equation}

\subsection{Mock signals of $m_g = 0 $}
\label{sec:MockGRSignal}

We first analyze \ref{item:U1}, \ref{item:L1}, \ref{item:L2} and \ref{item:P1} that are generated by assuming $m_g = 0 $.
The frequency-domain strains of \ref{item:U1} and \ref{item:P1} are generated using the \texttt{IMRPhenoPv2} template \cite{IMRPhenom_2011, IMRPhenom_2014}, a phenomenological waveform template calibrated against numerical-relativity simulations, using the \texttt{LALSimulation} library \cite{lalsuite}. 
The simulated unlensed signals contain the inspiral, merger, and ringdown phase. 
We then map \ref{item:U1} into \ref{item:L1} by multiplying the frequency-domain waveform of \ref{item:U1} by the amplification function \cref{eq:Amp_fn_def}.
\ref{item:L2} is also generated according to these procedures. 
When inferring \ref{item:L1} and \ref{item:L2}, we use the waveform model of \cref{eq:lensed_waveform} with the dephasing due to the massive gravitons included and infer $m_g$ along with $\vec{\theta}_{\rm BBH}$ and $\vec{\theta}_{\rm len}$. 
For \ref{item:U1} and \ref{item:P1}, we infer with the waveform model with $F(f; m_g) $ in \cref{eq:lensed_waveform} set to be $ 1 $ for all frequencies and $M_{\rm len}$ and $ y $ are removed from inference.

\begin{figure}[tp!]
  \centering
	\includegraphics[width=\columnwidth]{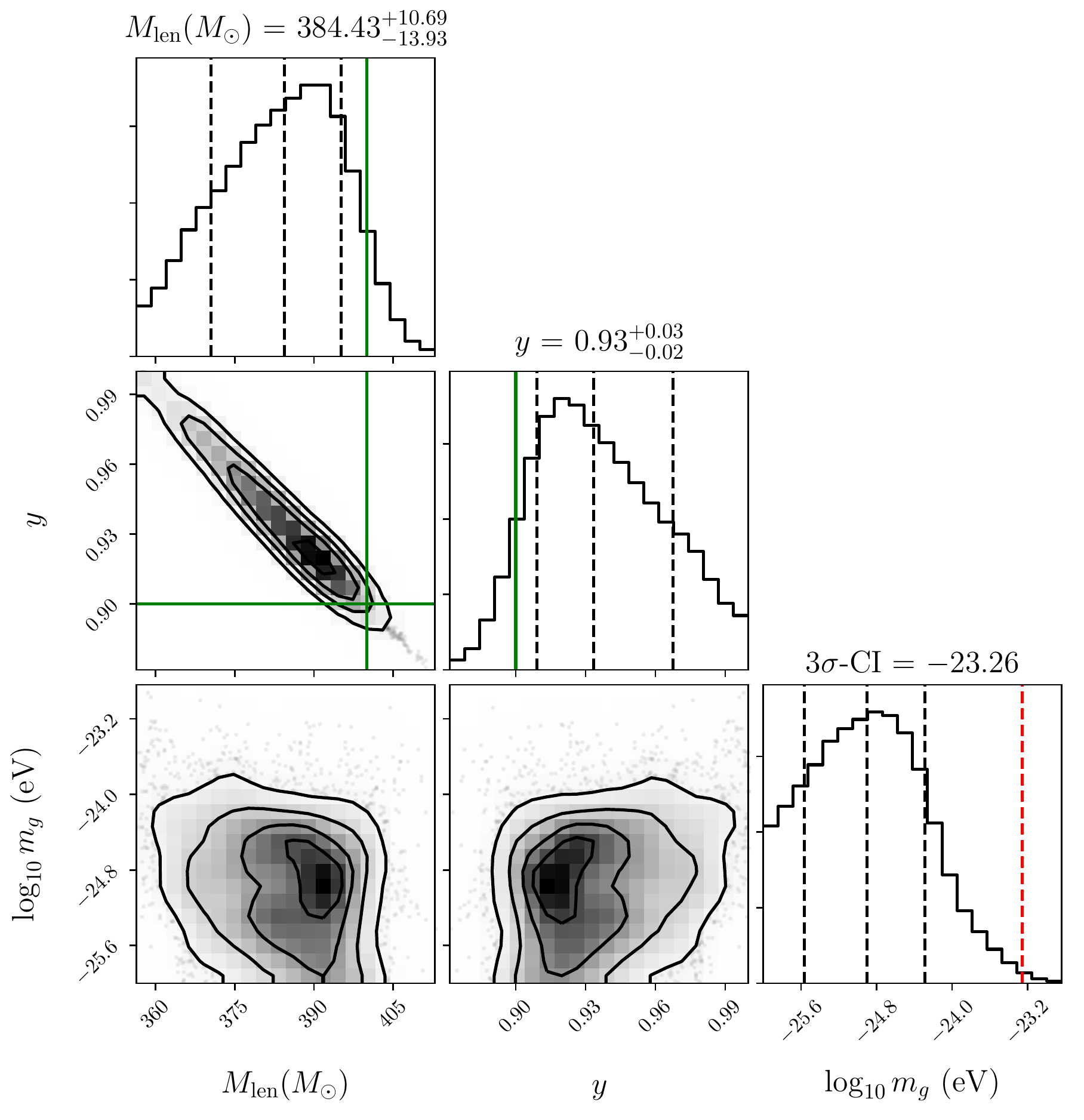}
	\caption{The corner plot shows the marginalized posterior of the redshifted lens mass $ M_{\rm len} $, $ y $ and $ \log m_g $ and their correlations, although we also infer the parameters of the source binary together as free parameters. 
	The posteriors are estimated from a mock lensed signal due to a GW150914-like binary lensed by a black hole of redshifted mass $M_{\rm len}$ of $ 400 M_{\odot}$ at $y=0.9$. 
	The green lines denote the injected values for $ M_{\rm len} $ and $ y $. 
	The red line on the marginalized posterior of $\log_{10} m_g $ denotes the $3\sigma$ confidence interval (CI) from the lower limit of the prior of $\log_{10} m_g$. 
	We conclude that we can bound the graviton mass while accurately measuring the lensing-related parameters.
 }
  \label{fig:corner_plot}
\end{figure}

The diagonal of \cref{fig:corner_plot} shows the posterior of redshifted lens mass $ M_{\rm len} $, $ y $ and $ \log m_g $  obtained from \ref{item:L1}. 
The off-diagonal plots show the two-dimensional posterior distributions among the variables. 
The green vertical lines mark the injected values.
The red vertical line marks the $3\sigma$ interval of the marginalized posterior of $\log_{10} m_g$ from $m_g = 10^{-26} \rm eV$. 
From \cref{fig:corner_plot}, we find that the posterior of $ \log_{10} m_g $ has no support for $ \log_{10} m_g > -23.2 $ because our measurement of GWs rules out the possibility of an excessive large $m_g$. 
From the posterior of $ M_{\rm len}$ and $ y $, we conclude that we can accurately estimate the lensing-related parameters while testing the graviton mass with lensing. 
Moreover, judging from \cref{fig:corner_plot} there are no strong correlations between the lensing-related parameters and $ m_g $.

\begin{figure}[tp!]
  \centering
	\includegraphics[width=\columnwidth]{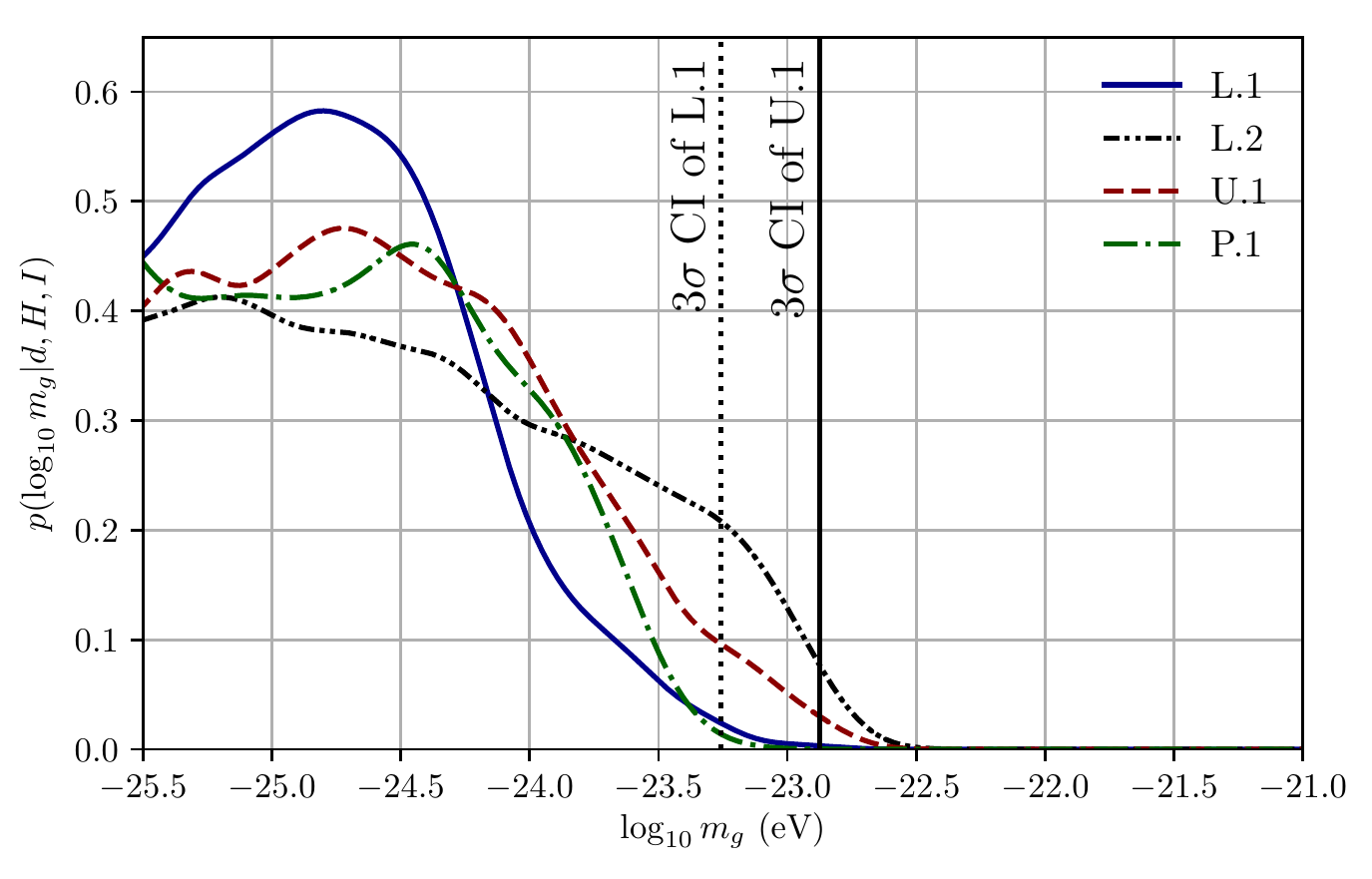}
	\caption{The marginalized posterior of $m_g$ in solid blue line is obtained from a lensed signal (\ref{item:L1}) due to a GW150914-like binary lensed by a black hole of redshifted mass of $ 400 M_{\odot}$ at $y=0.9$.
	To gauge the improvement of the constraint, we compare the result of a lensed signal with its unlensed counterpart (\ref{item:U1}, dashed red line). 
	To examine the role of the signal-to-noise ratio (SNR), we inject another lensed signal (\ref{item:L2}, in dot-dot-dashed line) that is identical to \ref{item:L1} except that the source binary is more farther away so that it has the same SNR of \ref{item:U1}.
	At last, lensing rate calculations suggest that the Advanced LIGO and Virgo may detect an IMBH-lensed signal per $\sim $ 7000 unlensed signals. 
	For a fair comparison, we also compare the lensed result with the best constraint likely to be obtained from 7000 simulated unlensed signals (\ref{item:P1}, dotted green line, see its generation in the main text). 
	The solid vertical line denotes the 3$\sigma$ CI of the posterior of \ref{item:U1} ($1.3 \times 10^{-23} \rm eV $) and the dashed vertical line denotes the 3$\sigma$-CI of \ref{item:L1} ($5.5 \times 10^{-24} \rm eV $). 
    All posteriors correspond to a similar 3-$\sigma$ interval, ranging from ranging from $ 3.3 \times 10^{-24} \rm eV $ to $1.3 \times 10^{-23} \rm eV $.
	}
  \label{fig:Constraint}
\end{figure}

We now compare the constraints on $m_g$ obtained from different nondispersive GW signals.
\cref{fig:Constraint} shows the posterior of $ \log_{10} m_g $ of \ref{item:L1} (solid blue line), its unlensed counterpart \ref{item:U1} (dashed red line), \ref{item:L2} (dashed dotted dotted black line) and \ref{item:P1} (dashed dotted green). 
We notice that all posteriors are in step-function shape because the measurement rules out large values for $m_g$, which would produce discernible effects on the waveform.
All posteriors correspond to a similar $3\sigma$ confidence interval (CI), ranging from $ 3.3 \times 10^{-24} \rm eV $ to $1.3 \times 10^{-23} \rm eV $. 
In particular, \ref{item:L1} yields a constraint (3$\sigma$ CI) on $ m_g$ of $ 5.5 \times 10^{-24} \rm eV $, slightly better than the constraint on $m_g$ by \ref{item:U1}, corresponding to $ 1.3 \times 10^{-23} \rm eV $. 
At the same SNR, we find that the $3\sigma$ CI of \ref{item:L2} is $ 1.3 \times 10^{-23} \rm eV $, almost the same as that of \ref{item:U1}.
Even with large SNR, \ref{item:P1} yields a constraint on $m_g$ of $\sim 3.3 \times 10^{-24} \rm eV $, slightly better than the constraint by all the other signals. 
These results conclude that lensing and increasing the SNR do not significantly improve the constraints. 

\subsection{Mock signals of $m_g = 10^{-22} \rm eV $}

\begin{figure}[tp!]
  \centering
	\includegraphics[width=\columnwidth]{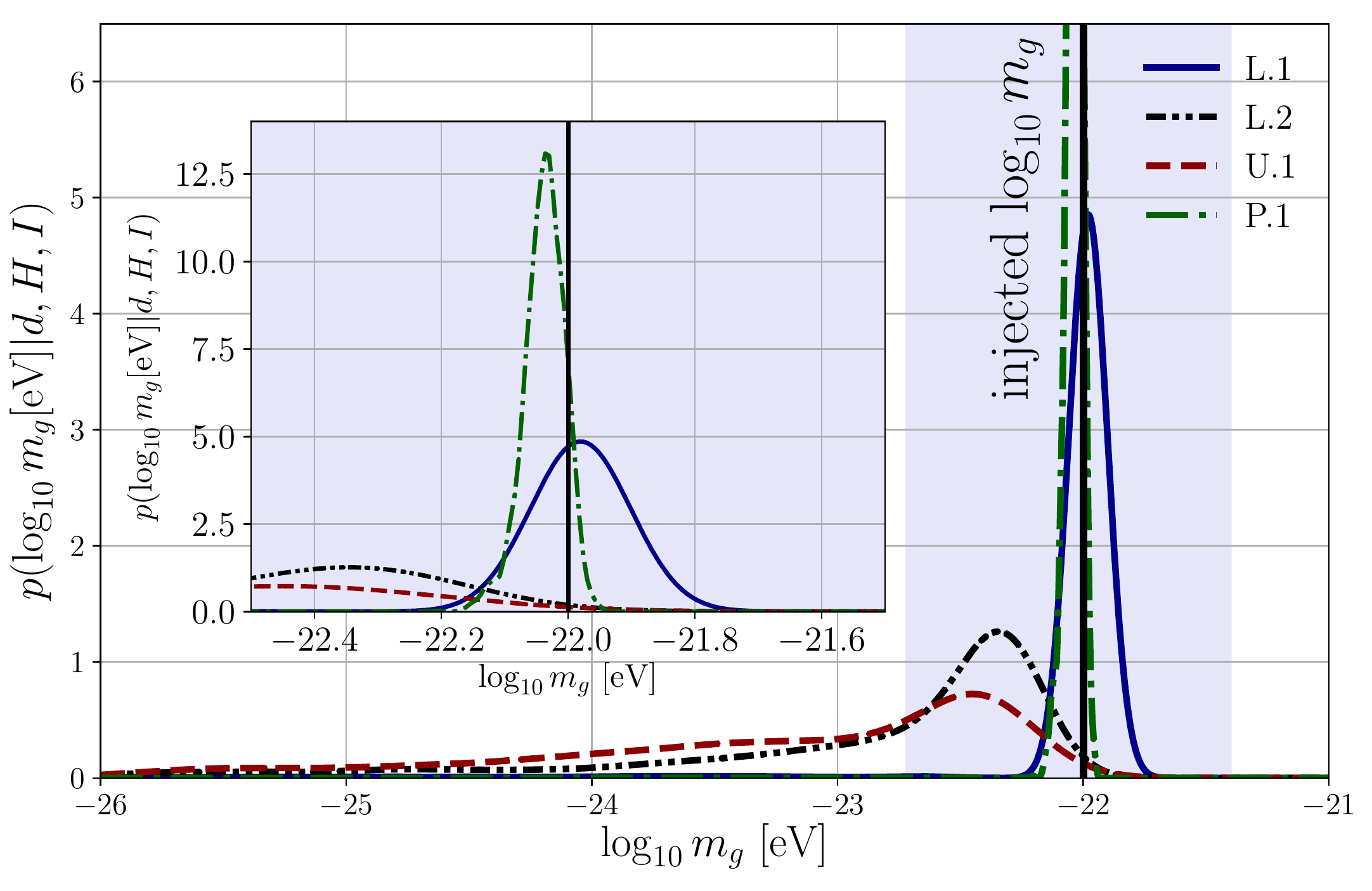}
	\caption{The marginalized posterior of $ m_g $ obtained from \ref{item:L1} (solid blue), \ref{item:U1} (dashed red), \ref{item:L2} (dashed-dotted-dotted black line) and \ref{item:P1} (dashed-dotted green).
	The embed figure shows the zoomed-in comparison of the posterior of \ref{item:L1} and \ref{item:P1}. 
	The shaped region denotes the 3$\sigma$ confidence interval of the posterior obtained from the lensed signal. 
	For all signals, we assume $m_g = 10^{-22} \mathrm{eV}$ (vertical line in black).
	The lensing geometry is the same as that considered in \cref{fig:Constraint} and \cref{sec:MockGRSignal}. 
	By comparing the posterior of \ref{item:L2} and \ref{item:U1}, we find that, even at the same SNR lensing modifications of waveform morphology contribute to improving measurement accuracy of $m_g$.  
	}
  \label{fig:Non_GR_Injection}
\end{figure}

On the other hand, we find that lensing can help to improve the measurement of $m_g$ from dispersive GWs. 
\cref{fig:Non_GR_Injection} shows the posterior of $\log_{10} m_g $ obtained from \ref{item:U1} (dashed red line), \ref{item:L1} (solid blue line), \ref{item:L2} (dot-dot-dashed line) and \ref{item:P1} (dot-dashed green line) that are generated by assuming $m_g = 10^{-22} \rm eV$ (solid vertical black line).
The shaded region illustrates the $3\sigma$ CI of the posterior of \ref{item:L1}. 
The embed figure shows the zoomed-in comparison of the posterior of \ref{item:L1} and \ref{item:P1}. 
We notice that, at the same SNR, lensing still improves the measurement accuracy of $m_g$ over its unlensed counterpart. 
The posterior of \ref{item:L2} shows more support for $m_g$ close to the injected $m_g$ than \ref{item:U1}, indicating that the the posterior of \ref{item:L2} is more accurate than that of \ref{item:U1}.
This is because lensing modulates the amplitude and phase of GWs, so that lensed GWs depend on $m_g$ more sensitively, increasing the detectability of dispersive GWs, as indicated by \cref{fig:match}. 
The posterior of \ref{item:L1} and \ref{item:P1} peaks at a $m_g$ closer to the injected $m_g$ because of larger SNR. 
Nevertheless, \ref{item:L1} still leads to significant improvement of the measurement accuracy of $m_g$ to an extent comparable to \ref{item:P1}. 
From the results of \cref{fig:Non_GR_Injection}, we find that both the lensing modifications of waveform morphology and amplification can contribute to the improved measurement accuracy of the graviton mass. 

\section{Conclusions} 
This paper studies the lensing pattern by a point-mass lens of GWs with an isotropic dispersion relation due to the massive gravitons. 
Although the graviton mass close to the existing constraints leads to no significant effects on the lensing amplification, we find that lensing modifies the waveform morphology of dispersive GWs, making the morphology changes more sensitively with the graviton mass, which helps to improve the measurement of the graviton mass. 
The improvement can also be further enhanced by the increase of signal-to-noise ratio due to lensing. 
By detecting a lensed gravitational-wave signal, we can measure the graviton mass with an accuracy comparable with the combined measurement across $\mathcal{O}(10^3)$ unlensed signals. 
Our work lays the foundation for measuring the graviton mass in the era of detectable lensed GWs, making the existing analyses that focus primarily on unlensed signals more complete. 

Other than the improvement of measurement accuracy of the graviton mass, our method enjoys several advantages. 
First, compared to other proposed methods of testing the speed of GWs by observing lensing \cite{Collett_Bacon_2017, Fan_2017}, our approach requires no observation of the electromagnetic counterpart(s) of a given event. 
Therefore, our method is more stand-alone and is easier to be performed. 
Second, our method is independent of the nature of the source binaries.
Although in this paper, we focused on GWs generated by binary black holes, our method can be straightforwardly applied to other types of coalescence, such as binary neutron star coalescence \cite{Lensing_GW_BNS}.
This flexibility greatly extends the scope of graviton-mass measurement. 
Lastly, our method makes the test of graviton mass more complete. 
While the far-field propagation of GWs \cite{Will_2012, Ajith_2010} and near-field behavior of black holes \cite{Chung_2019} have been proposed to constrain the mass of graviton, our test bridges the intermediate region between these two tests.
Along with other tests of general relativity via observing the lensing of GWs (such as \cite{Lensing_TGR}), our test demonstrates the strong potential to understand the nature of space-time via observing gravitational-wave lensing.

In this work, we ignore the effects of (i) the change of polarization of GWs due to lensing \cite{Lensing_polarization}; (ii) the change of the behavior of the source compact binary due to massive graviton, as is the case in \cite{LIGO_07, LIGO_11}; and (iii) the change of the gravitational field around the lens by the graviton mass. 
Also, our study focusing on the case of point-mass lens. 
These ignored effects and the lensing of dispersive GWs of other lens types remain fully explored.
If we include these effects in our measurement, the accuracy can be further enhanced. 

In the future, we plan to extend our studies to other types of lenses, which may help further improve the measurement accuracy. 
Our study has thus far focused on the point-mass lens, such as intermediate-mass black holes, which leads to microlensing.
In reality, it may be very rare for gravitational waves to be lensed by an intermediate-mass black hole of $\sim 400 M_{\odot}$. 
Moreover, the population properties and lensing rates of intermediate-mass black holes are uncertain. 
On the other hand, strong lensing due to different types of lenses, such as galaxies or galaxy clusters \cite{galaxy_strong_lensing_01}, are expected to be more common, roughly 1 per $\sim 600$ unlensed events at the design sensitivity of LIGO and Virgo \cite{Ken_2016}. 
Upon strong lensing, a GW signal may split into multiple images whose properties, such as image position and the arrival time differences, may depend on the graviton mass even more sensitively than the diffraction pattern considered in this work \cite{Baker, private_communication}. 
To extend our test to strong lensing, we need to study the strong lensing of dispersive GWs by lenses with structures, such as galactic lenses, singular isothermal sphere, and other possible extended mass distribution \cite{lensing_mass_distribution}. 
We would also like to investigate the performance of our test for the detection by proposed space-based detectors, such as the Laser Interferometer Space Antenna \cite{LISA}, which are capable of exquisite phase measurement and much better constraints.
Therefore, in the future, we can measure the graviton mass with unparalleled accuracy by observing lensed gravitational-wave signals. 

\section*{Acknowledgements}

The authors are indebted to valuable discussion among the lensing working group of LIGO.
A.K.-W.C. would like to acknowledge Patrick C.K. Cheong, Srashti Goyal and Shasvath Kapadia for stimulating discussions, Jose Maria Ezquiaga Bravo, Mark H.Y. Cheung, Otto A. Hannuksela, Alvin K.Y. Li and Ignacio Magana for their comments on the manuscript and relevant presentations and Robin S.H. Yuen for his advice about computer programming.
A.K.-W.C. was supported by the Hong Kong Scholarship for Excellence Scheme (HKSES). 
The work described in this paper was partially supported by grants from the Research Grants Council of the Hong Kong (Project No. CUHK 24304317 and CUHK 14306218), The Croucher Foundation of Hong Kong, and the Research Committee of the Chinese University of Hong Kong.
This manuscript carries a report number of KCL-PH-TH 2021/41 and LIGO Document number of P2100192-v2. 

This research has made use of data, software and/or web tools obtained from the GW Open Science Center (https://www.gw-openscience.org), a service of LIGO Laboratory, the LIGO Scientific Collaboration and the Virgo Collaboration. LIGO is funded by the U.S. National Science Foundation. Virgo is funded by the French Centre National de Recherche Scientifique (CNRS), the Italian Istituto Nazionale della Fisica Nucleare (INFN) and the Dutch Nikhef, with contributions by Polish and Hungarian institutes.

\bibliography{bibtex}

\begin{thebibliography}{10}

\bibitem{Virgo_noise_curve}
{\url{https://dcc.ligo.org/LIGO-P1200087-v42/public}}.

\bibitem{LIGO_noise_curve}
{\url{https://dcc.ligo.org/LIGO-T2000012/public}}.

\bibitem{IMRPhenom_2011}
P.~Ajith, M.~Hannam, S.~Husa, Y.~Chen, B.~Br\"ugmann, N.~Dorband, D.~M\"uller,
  F.~Ohme, D.~Pollney, C.~Reisswig, L.~Santamar\'{\i}a, and J.~Seiler.
\newblock Inspiral-merger-ringdown waveforms for black-hole binaries with
  nonprecessing spins.
\newblock {\em Phys. Rev. Lett.}, 106:241101, Jun 2011.

\bibitem{LIGO_08}
{B. P. Abbott \textit{et al.} }.
\newblock Binary black hole mergers in the first advanced ligo observing run.
\newblock {\em Phys. Rev. X}, 6:041015, Oct 2016.

\bibitem{LIGO_02}
{B. P. Abbott \textit{et al.} }.
\newblock Gw151226: Observation of gravitational waves from a 22-solar-mass
  binary black hole coalescence.
\newblock {\em Phys. Rev. Lett.}, 116:241103, Jun 2016.

\bibitem{LIGO_01}
{B. P. Abbott \textit{et al.} }.
\newblock Observation of gravitational waves from a binary black hole merger.
\newblock {\em Phys. Rev. Lett.}, 116:061102, Feb 2016.

\bibitem{LIGO_14}
{B. P. Abbott \textit{et al.} }.
\newblock {Properties of the Binary Black Hole Merger GW150914}.
\newblock {\em prl}, 116(24):241102, June 2016.

\bibitem{LIGO_07}
{B. P. Abbott \textit{et al.} }.
\newblock Tests of general relativity with gw150914.
\newblock {\em Phys. Rev. Lett.}, 116:221101, May 2016.

\bibitem{LIGO_03}
{B. P. Abbott \textit{et al.} }.
\newblock Gw170104: Observation of a 50-solar-mass binary black hole
  coalescence at redshift 0.2.
\newblock {\em Phys. Rev. Lett.}, 118:221101, Jun 2017.

\bibitem{LIGO_04}
{B. P. Abbott \textit{et al.} }.
\newblock Gw170608: Observation of a 19 solar-mass binary black hole
  coalescence.
\newblock {\em The Astrophysical Journal Letters}, 851(2):L35, 2017.

\bibitem{LIGO_05}
{B. P. Abbott \textit{et al.} }.
\newblock Gw170814: A three-detector observation of gravitational waves from a
  binary black hole coalescence.
\newblock {\em Phys. Rev. Lett.}, 119:141101, Oct 2017.

\bibitem{LIGO_06}
{B. P. Abbott \textit{et al.} }.
\newblock Gw170817: Observation of gravitational waves from a binary neutron
  star inspiral.
\newblock {\em Phys. Rev. Lett.}, 119:161101, Oct 2017.

\bibitem{LIGO_09}
{B. P. Abbott \textit{et al.} }.
\newblock {Prospects for Observing and Localizing Gravitational-Wave Transients
  with Advanced LIGO, Advanced Virgo and KAGRA}.
\newblock {\em Living Rev. Rel.}, 21:3, 2018.
\newblock [Living Rev. Rel.19,1(2016)].

\bibitem{LIGO_10}
{B. P. Abbott \textit{et al.} }.
\newblock Gwtc-1: A gravitational-wave transient catalog of compact binary
  mergers observed by ligo and virgo during the first and second observing
  runs.
\newblock {\em Phys. Rev. X}, 9:031040, Sep 2019.

\bibitem{LIGO_11}
{B. P. Abbott \textit{et al.} }.
\newblock Tests of general relativity with the binary black hole signals from
  the ligo-virgo catalog gwtc-1.
\newblock {\em Phys. Rev. D}, 100:104036, Nov 2019.

\bibitem{LIGO_12}
{B. P. Abbott \textit{et al.} }.
\newblock {GW190412: Observation of a Binary-Black-Hole Coalescence with
  Asymmetric Masses}.
\newblock {\em arXiv e-prints}, page arXiv:2004.08342, April 2020.

\bibitem{LIGO_13}
{B. P. Abbott \textit{et al.} }.
\newblock {GW}190814: Gravitational waves from the coalescence of a 23 solar
  mass black hole with a 2.6 solar mass compact object.
\newblock {\em The Astrophysical Journal}, 896(2):L44, jun 2020.

\bibitem{LIGO_16}
{B. P. Abbott \textit{et al.} }.
\newblock {GWTC-2: Compact Binary Coalescences Observed by LIGO and Virgo
  During the First Half of the Third Observing Run}.
\newblock {\em arXiv e-prints}, page arXiv:2010.14527, October 2020.

\bibitem{LIGO_17}
{B. P. Abbott \textit{et al.} }.
\newblock {Population Properties of Compact Objects from the Second LIGO-Virgo
  Gravitational-Wave Transient Catalog}.
\newblock {\em arXiv e-prints}, page arXiv:2010.14533, October 2020.

\bibitem{LIGO_15}
{B. P. Abbott \textit{et al.} }.
\newblock {Tests of General Relativity with Binary Black Holes from the second
  LIGO-Virgo Gravitational-Wave Transient Catalog}.
\newblock {\em arXiv e-prints}, page arXiv:2010.14529, October 2020.

\bibitem{O3a_lensing_paper}
{B. P. Abbott \textit{et al.} }.
\newblock {Search for lensing signatures in the gravitational-wave observations
  from the first half of LIGO-Virgo's third observing run}.
\newblock {\em arXiv e-prints}, page arXiv:2105.06384, May 2021.

\bibitem{LIGO_detection_rate}
Vishal Baibhav, Emanuele Berti, Davide Gerosa, Michela Mapelli, Nicola
  Giacobbo, Yann Bouffanais, and Ugo~N. Di~Carlo.
\newblock Gravitational-wave detection rates for compact binaries formed in
  isolation: Ligo/virgo o3 and beyond.
\newblock {\em Phys. Rev. D}, 100:064060, Sep 2019.

\bibitem{Baker}
Tessa Baker and Mark Trodden.
\newblock Multimessenger time delays from lensed gravitational waves.
\newblock {\em Phys. Rev. D}, 95:063512, Mar 2017.

\bibitem{Lensing_GW_02}
Haugan Mark~P Bontz, Robert~J.
\newblock {A diffraction limit on the gravitational lens effect}.
\newblock {\em Astrophysics and Space Science}, August 1981.

\bibitem{Sathyaprakash_2009}
Alessandra Buonanno, Bala~R. Iyer, Evan Ochsner, Yi~Pan, and B.~S.
  Sathyaprakash.
\newblock Comparison of post-newtonian templates for compact binary inspiral
  signals in gravitational-wave detectors.
\newblock {\em Phys. Rev. D}, 80:084043, Oct 2009.

\bibitem{Chung_2019}
Adrian Ka-Wai Chung and Tjonnie Guang~Feng Li.
\newblock Phenomenological inclusion of alternative dispersion relations to the
  teukolsky equation and its application to bounding the graviton mass with
  gravitational-wave measurements.
\newblock {\em Phys. Rev. D}, 99:124023, Jun 2019.

\bibitem{Collett_Bacon_2017}
T.~E. {Collett} and D.~{Bacon}.
\newblock {Testing the Speed of Gravitational Waves over Cosmological Distances
  with Strong Gravitational Lensing}.
\newblock {\em Physical Review Letters}, 118(9):091101, March 2017.

\bibitem{Cremonese2021}
Paolo {Cremonese}, Jose {Mar{\'\i}a Ezquiaga}, and Vincenzo {Salzano}.
\newblock {Breaking the mass-sheet degeneracy with gravitational wave
  interference in lensed events}.
\newblock {\em arXiv e-prints}, page arXiv:2104.07055, April 2021.

\bibitem{MG_review_01}
Claudia {de Rham}.
\newblock {Massive Gravity}.
\newblock {\em Living Reviews in Relativity}, 17(1):7, December 2014.

\bibitem{m_g_bound_review_01}
Claudia de~Rham, J.~Tate Deskins, Andrew~J. Tolley, and Shuang-Yong Zhou.
\newblock Graviton mass bounds.
\newblock {\em Rev. Mod. Phys.}, 89:025004, May 2017.

\bibitem{Lensing_GW_03}
S.~{Deguchi} and W.~D. {Watson}.
\newblock {Diffraction in Gravitational Lensing for Compact Objects of Low
  Mass}.
\newblock {\em \apj}, 307:30, August 1986.

\bibitem{Schneider_01}
P.~Schneider J. Ehlers~E.E. Falco.
\newblock {\em Gravitational Lenses}.
\newblock Springer-Verlag Berlin Heidelberg, 1992.

\bibitem{Fan_2017}
X.-L. {Fan}, K.~{Liao}, M.~{Biesiada}, A.~{Pi{\'o}rkowska-Kurpas}, and Z.-H.
  {Zhu}.
\newblock {Speed of Gravitational Waves from Strongly Lensed Gravitational
  Waves and Electromagnetic Signals}.
\newblock {\em Physical Review Letters}, 118(9):091102, March 2017.

\bibitem{Inner_product}
\'Eanna~\'E. Flanagan and Scott~A. Hughes.
\newblock Measuring gravitational waves from binary black hole coalescences. i.
  signal to noise for inspiral, merger, and ringdown.
\newblock {\em Phys. Rev. D}, 57:4535--4565, Apr 1998.

\bibitem{Lensing_TGR}
Srashti {Goyal}, K.~{Haris}, Ajit~Kumar {Mehta}, and Parameswaran {Ajith}.
\newblock {Testing the nature of gravitational-wave polarizations using
  strongly lensed signals}.
\newblock {\em arXiv e-prints}, page arXiv:2008.07060, August 2020.

\bibitem{private_communication}
Srashti Goyal and Shasvath Kapadia.
\newblock private communication, May 2021.

\bibitem{Lensing_O1_O2}
O.~A. {Hannuksela}, K.~{Haris}, K.~K.~Y. {Ng}, S.~{Kumar}, A.~K. {Mehta},
  D.~{Keitel}, T.~G.~F. {Li}, and P.~{Ajith}.
\newblock {Search for Gravitational Lensing Signatures in LIGO-Virgo Binary
  Black Hole Events}.
\newblock {\em apjl}, 874(1):L2, March 2019.

\bibitem{Hubble_constant}
Otto~A. {Hannuksela}, Thomas~E. {Collett}, Mesut {{\c{C}}al{\i}{\textcommabelow
  s}kan}, and Tjonnie G.~F. {Li}.
\newblock {Localizing merging black holes with sub-arcsecond precision using
  gravitational-wave lensing}.
\newblock {\em mnras}, 498(3):3395--3402, August 2020.

\bibitem{Yagi_2015}
D.~{Hansen}, N.~{Yunes}, and K.~{Yagi}.
\newblock {Projected constraints on Lorentz-violating gravity with
  gravitational waves}.
\newblock {\em \prd}, 91(8):082003, April 2015.

\bibitem{Lensing_polarization}
Shaoqi Hou, Xi-Long Fan, and Zong-Hong Zhu.
\newblock Gravitational lensing of gravitational waves: Rotation of
  polarization plane.
\newblock {\em Phys. Rev. D}, 100:064028, Sep 2019.

\bibitem{Ken_2016}
Tjonnie G. F.~Li Ken K. Y.~Ng, Kaze W. K.~Wong and Tom Broadhurst.
\newblock Precise ligo lensing rate predictions for binary black holes.
\newblock {\em \prd}, 2017.

\bibitem{Ajith_2010}
D.~Keppel and P.~Ajith.
\newblock Constraining the mass of the graviton using coalescing black-hole
  binaries.
\newblock {\em Phys. Rev. D}, 82:122001, Dec 2010.

\bibitem{Lensing_IMBH_01}
Kwun-Hang {Lai}, Otto~A. {Hannuksela}, Antonio {Herrera-Mart{\'\i}n}, Jose~M.
  {Diego}, Tom {Broadhurst}, and Tjonnie G.~F. {Li}.
\newblock {Discovering intermediate-mass black hole lenses through
  gravitational wave lensing}.
\newblock {\em \prd}, 98(8):083005, October 2018.

\bibitem{lalsuite}
{LIGO Scientific Collaboration}.
\newblock {LIGO} {A}lgorithm {L}ibrary - {LALS}uite.
\newblock free software (GPL), 2018.

\bibitem{Lensing_phase_shift_01}
Jose {Mar{\'\i}a Ezquiaga}, Daniel~E. {Holz}, Wayne {Hu}, Macarena {Lagos}, and
  Robert~M. {Wald}.
\newblock {Phase effects from strong gravitational lensing of gravitational
  waves}.
\newblock {\em arXiv e-prints}, page arXiv:2008.12814, August 2020.

\bibitem{Lensing_birefringence}
Jose {Mar{\'\i}a Ezquiaga} and Miguel {Zumalac{\'a}rregui}.
\newblock {Gravitational wave lensing beyond general relativity: birefringence,
  echoes and shadows}.
\newblock {\em arXiv e-prints}, page arXiv:2009.12187, September 2020.

\bibitem{Will_2012}
S.~{Mirshekari}, N.~{Yunes}, and C.~M. {Will}.
\newblock {Constraining Lorentz-violating, modified dispersion relations with
  gravitational waves}.
\newblock {\em \prd}, 85(2):024041, January 2012.

\bibitem{LIGO_sensitivity}
C.~J. {Moore}, R.~H. {Cole}, and C.~P.~L. {Berry}.
\newblock {Gravitational-wave sensitivity curves}.
\newblock {\em Classical and Quantum Gravity}, 32:015014, Jan 2015.

\bibitem{Lensing_LIGO_01}
Suvodip {Mukherjee}, Tom {Broadhurst}, Jose~M. {Diego}, Joseph {Silk}, and
  George~F. {Smoot}.
\newblock {Inferring the lensing rate of LIGO-Virgo sources from the stochastic
  gravitational wave background}.
\newblock {\em mnras}, 501(2):2451--2466, February 2021.

\bibitem{Suvodip01}
Suvodip {Mukherjee}, Benjamin~D. {Wandelt}, and Joseph {Silk}.
\newblock {Multimessenger tests of gravity with weakly lensed gravitational
  waves}.
\newblock {\em \prd}, 101(10):103509, May 2020.

\bibitem{Suvodip02}
Suvodip {Mukherjee}, Benjamin~D. {Wandelt}, and Joseph {Silk}.
\newblock {Probing the theory of gravity with gravitational lensing of
  gravitational waves and galaxy surveys}.
\newblock {\em mnras}, 494(2):1956--1970, May 2020.

\bibitem{Nakamura_02}
Takahiro~T. Nakamura and Shuji Deguchi.
\newblock Wave optics in gravitational lensing.
\newblock {\em Progress of Theoretical Physics Supplement}, 133:137--153, 1999.

\bibitem{Lensing_GW_01}
Hans~C. Ohanian.
\newblock {On the focusing of gravitational radiation}.
\newblock {\em International Journal of Theoretical Physics}, June 1974.

\bibitem{lensingGW}
G.~{Pagano}, O.~A. {Hannuksela}, and T.~G.~F. {Li}.
\newblock {LENSINGGW: a PYTHON package for lensing of gravitational waves}.
\newblock {\em aap}, 643:A167, November 2020.

\bibitem{IMRPhenom_2014}
Yi~Pan, Alessandra Buonanno, Andrea Taracchini, Lawrence~E. Kidder, Abdul~H.
  Mrou\'e, Harald~P. Pfeiffer, Mark~A. Scheel, and B\'ela Szil\'agyi.
\newblock Inspiral-merger-ringdown waveforms of spinning, precessing black-hole
  binaries in the effective-one-body formalism.
\newblock {\em Phys. Rev. D}, 89:084006, Apr 2014.

\bibitem{Lensing_GW_BNS}
Peter T.~H. {Pang}, Otto~A. {Hannuksela}, Tim {Dietrich}, Giulia {Pagano}, and
  Ian~W. {Harry}.
\newblock {Lensed or not lensed: determining lensing magnifications for binary
  neutron star mergers from a single detection}.
\newblock {\em mnras}, 495(4):3740--3750, May 2020.

\bibitem{Yunes_2018}
Scott {Perkins} and Nicol{\'a}s {Yunes}.
\newblock {Probing screening and the graviton mass with gravitational waves}.
\newblock {\em Classical and Quantum Gravity}, 36(5):055013, March 2019.

\bibitem{LISA}
Travis Robson, Neil~J Cornish, and Chang Liug.
\newblock The construction and use of lisa sensitivity curves.
\newblock {\em Classical and Quantum Gravity}, 36(10):105011, 2019.

\bibitem{Schneider_02}
P.~Schneider.
\newblock {\em Introduction to Gravitational Lensing and Cosmology}, pages
  1--89.
\newblock Springer Berlin Heidelberg, Berlin, Heidelberg, 2006.

\bibitem{galaxy_strong_lensing_01}
Graham~P. Smith, Christopher Berry, Matteo Bianconi, Will~M. Farr, Mathilde
  Jauzac, Richard Massey, Johan Richard, Andrew Robertson, Keren Sharon,
  Alberto Vecchio, and et~al.
\newblock Strong-lensing of gravitational waves by galaxy clusters.
\newblock {\em Proceedings of the International Astronomical Union},
  13(S338):98–102, 2017.

\bibitem{Takahashi_2017}
R.~{Takahashi}.
\newblock {Arrival Time Differences between Gravitational Waves and
  Electromagnetic Signals due to Gravitational Lensing}.
\newblock {\em \apj}, 835:103, January 2017.

\bibitem{Lensing_IMBH_02}
Ryuichi {Takahashi} and Takashi {Nakamura}.
\newblock {Wave Effects in the Gravitational Lensing of Gravitational Waves
  from Chirping Binaries}.
\newblock {\em \apj}, 595(2):1039--1051, October 2003.

\bibitem{Takahashi_2003}
Ryuichi Takahashi and Takashi Nakamura.
\newblock Wave effects in the gravitational lensing of gravitational waves from
  chirping binaries.
\newblock {\em The Astrophysical Journal}, 595(2):1039, 2003.

\bibitem{lensing_mass_distribution}
Slava~G. {Turyshev} and Viktor~T. {Toth}.
\newblock {Gravitational lensing by an extended mass distribution}.
\newblock {\em arXiv e-prints}, page arXiv:2106.06696, June 2021.

\bibitem{m_g_bound_review_02}
Clifford~M. Will.
\newblock Bounding the mass of the graviton using gravitational-wave
  observations of inspiralling compact binaries.
\newblock {\em Phys. Rev. D}, 57:2061--2068, Feb 1998.

\end{thebibliography}
\bibliographystyle{plain}

\end{document}